\documentstyle[preprint,aps]{revtex}
\begin{document}
\title{
 A Lanczos  algorithm for linear response}
\author{C.W. Johnson$^{(a)}$, G.F. Bertsch$^{(b)}$\footnote{
E-mail: bertsch@phys.washington.edu}, and W.D. Hazelton$^{(b)}$}

\address{$^{(a)}$Department of Physics and Astronomy\\
Louisiana State University, Baton Rouge, LA 70803-4001\\
$^{(b)}$Dept. of Physics and Inst. Nuclear Theory\\
Box 351560\\
University of Washington
Seattle, WA 98915\\}

\maketitle

\begin{abstract}
      An iterative algorithm is presented for
solving the RPA equations of linear response.
The method optimally computes the energy-weighted moments of the 
strength function, allowing one to match the computational
effort to the intrinsic accuracy of the basic
mean-field approximation, avoiding the problem of solving very 
large matrices. For  local interactions, the computational 
effort for the method scales with the number
of particles $N_p$ as $O(N_p^3)$.
\end{abstract}

\section{Introduction}
   In a number of branches of physics, mean field theory gives
a remarkably effective approximation to the ground state. Similarly,
for the response of the system to small perturbations,  time-dependent 
mean field theory is a useful extension.
This is the experience in nuclear physics\cite{ri80,bo81},
atomic and molecular physics\cite{ya96,za80,ru96,ya93,lu94,ja96}
and condensed matter physics\cite{bl95,qu93}.  
There are of course intrinsic
limitations to these approximations, but equally pressing is the
large computational resources required for calculations of 
systems of interest. This is our motivation to look for algorithms
that better match the computational effort  to the intrinsic limits
of the approximation.

We take our inspiration from the Lanczos algorithm \cite{cu85}, 
which is best known in many-body physics for 
extracting low-lying eigenstates of very large 
Hamiltonian matrices\cite{ha80,wh77}.  When dealing with large
spaces, the computational question often comes down to the number 
of times the Hamiltonian operates on 
a state vector.  Depending on the Hamiltonian and the starting
vector, the Lanczos algorithm is able to extract an
accurate ground state vector in a basis of $10^{5-6}$ states 
with a few  hundred Hamiltonian operations.  
The algorithm may be viewed as a numerically stable technique\cite{pa72}
to compute
moments of the Hamiltonian with respect to some 
initial state $\Psi_0$, that is, 
$\mu_k \equiv 
\left \langle \Psi_0 \right | \hat{H}^k \left | \Psi_0  \right \rangle.$
For large $k$, $\mu_k$ is dominated by the extremal eigenvalues \cite{wh80},
which are thus available for recovery.

The Lanczos algorithm has also been applied to many other topics in atomic, 
molecular, solid state, and nuclear physics, including computation of 
the S-matrix \cite{ya89}, time-evolution of wave packets \cite{pa86}, 
level densities 
\cite{de89}, and the continued-fraction expansion of the
resolvant or Green's function \cite{resolvant}.  
Particularly relevant to us is the application to strength functions.  
The strength function $S$ for an operator $\hat Q$ on a state $i$
is defined
\begin{equation}
S(E) \equiv \sum_f \delta(E-E_f+E_i) \left | 
\left \langle f \right | \hat{Q} \left | i \right \rangle \right |^2.
\end{equation}
A powerful technique to calculate the strength function,  
successfully applied to the nuclear shell model \cite{wh77,ca90},
uses the Lanczos algorithm with a starting vector
$\left | \Psi_0 \right \rangle = \hat{Q} \left | i \right \rangle/\langle
i|\hat Q^2|i\rangle^{1/2}$.
The Lanczos algorithm implicitly computes the moments 
$$
M_k=\int\,dE (E-E_i)^k S(E)
$$ 
of the strength function.  
After a few tens of iterations one 
can accurately reconstruct the distribution of the exact strength 
function. 

In many cases, however, the matrix elements of the operator $\hat Q$ are
sensitive to correlations in the ground state, and then the size of the
wave function basis in the straightforward Hamiltonian approach 
becomes problematic.  In this situation, the time-dependent mean-field
theory offers a reasonable compromise.
The small amplitude theory, the RPA or linear response, can be cast 
into a matrix form in a particle-hole basis.  However,
the RPA matrix is not symmetric as required by the Lanczos algorithm.
The matrix equation is commonly written as
\def\be{\begin{equation}}
\def\ee{\end{equation}}
\be
\left(\matrix{ {\bf A} & {\bf B}\cr 
-{\bf B} &-{\bf A}\cr}\right)\left(\matrix{\vec{x}\cr \vec{y}}
\right)=\omega\left(\matrix{\vec{x}\cr \vec{y}}\right)
\label{RPA}
\ee
where ${\bf A}$ and ${\bf B}$ are particle-hole Hamiltonian matrices, 
$\omega$ is
the eigenfrequency, and $\vec{x}$ and $\vec{y}$ 
are the vectors of positive- and negative-frequency 
particle-hole
amplitudes, respectively.  
An important property of the RPA equation is that eigenvectors 
come in conjugate pairs: in equation (\ref{RPA}) 
$(\vec{y},\vec{x})$ is also an eigenvector with 
eigenfrequency $-\omega$.  
For the linear response, 
the matrix element between the RPA ground
state $|0\rangle$ and an excited state $|\omega\rangle$ may be expressed
as
\be
\langle\omega|\hat Q|0\rangle=\left(\vec{q} ,\vec{q} \right) \cdot 
\left(\matrix{ \vec{x}\cr \vec{y}}\right)
\ee
where $\vec{q}$ is the vector of particle-hole matrix elements and the 
vector $(\vec{x},\vec{y})$ is normalized as
$
1=\vec{x}\cdot \vec{x} - \vec{y}\cdot \vec{y}.
$

There are a number of ways to introduce a Lanczos-type algorithm for 
the RPA matrix.  The method we describe here has the advantages that it
preserves the form eq.(\ref{RPA}) of the RPA matrix and it produces strength
functions that respect sum rules.  We seek a new basis of vectors
$|Z_i\rangle := (\vec X_i, \vec Y_i) $ where the matrices of
column vectors ${\bf U} := ( \vec{X}_1, \vec{X}_2, \vec{X}_3, \ldots )$ 
and  ${\bf V} := ( \vec{Y}_1, \vec{Y}_2, \vec{Y}_3, \ldots )$ 
transform the RPA matrix as
\begin{equation}
\left ( \matrix{ {\bf U}^T & -{\bf V}^T \cr -{\bf V}^T & {\bf U}^T  
} \right ) 
\left ( \matrix{ {\bf A} & {\bf B} \cr -{\bf B} & -{\bf A}  
} \right ) 
\left ( \matrix{ {\bf U} & {\bf V} \cr {\bf V} & {\bf U} 
} \right ) 
=\\
\left ( \matrix{ {\bf A}^\prime & {\bf B}^\prime \cr -{\bf B}^\prime & 
-{\bf A}^\prime  
} \right ) \label{transform}
\end{equation}
where the transformed matrices $\bf A'$ and $\bf B'$ 
are now tridiagonal: 
\begin{eqnarray}
{ \bf A}^\prime = \left(\matrix{e_1&a_1&0&\cr a_1&e_2 &a_2&\cr 0&a_2&e_3&\cr
&&&\ddots}\right),  
{\bf B}^\prime =\left(\matrix{d_1&b_1&0&\cr
b_1& d_2& b_2&\cr 0& b_2& d_3& \cr &&&\ddots}\right)
\label{TransformedMatrix}
\end{eqnarray}

The Lanczos basis vectors and 
matrix elements are generated iteratively as follows. 
 Suppose we have the vectors $|Z_1\rangle,...,|Z_n\rangle$
already computed, together with
the transformed matrix up to $e_{n-1},d_{n-1},a_{n-1}$ and $b_{n-1}$.  
The iteration starts
by applying the RPA matrix in eqn. (\ref{RPA}) to the vector $|Z_n\rangle$,
\be
|Z_t\rangle= \left(\matrix{\vec{X}_t\cr \vec{Y}_t}\right) =
\left(\matrix{{ \bf A} \vec{X}_n + {\bf B} \vec{Y}_n \cr 
-{\bf B} \vec{X}_n -{\bf A} \vec{Y}_n}\right)
\ee
The diagonal elements $e_n$ and $d_n$ are now easily computed:
\begin{eqnarray}
e_n = \vec{X}_t\cdot \vec{X}_n - \vec{Y}_t\cdot \vec{Y}_n 
\nonumber \\
d_n = \vec{X}_t\cdot \vec{Y}_n - \vec{Y}_t\cdot \vec{X}_n.
\end{eqnarray}
We next project out $|Z_t'\rangle$, the component of $|Z_t\rangle$ that is 
orthogonal to the space $|Z_1\rangle,...,|Z_n\rangle$.
This can be done conveniently by using the matrix elements
in (\ref{TransformedMatrix}) that have already been calculated,
\be
|Z_t'\rangle=\left(\matrix{\vec{X}_t'\cr \vec{Y}_t'}\right)
=\\
\left(\matrix{\vec{X}_t-e_{n}\vec{X}_{n}+
d_{n}\vec{Y}_{n}-a_{n-1}\vec{X}_{n-1}+b_{n-1}\vec{Y}_{n-1}\cr 
\vec{Y}_t -d_{n} \vec{X}_{n}
+e_{n} \vec{Y}_{n}-b_{n-1} \vec{X}_{n-1}+a_{n-1}\vec{Y}_{n-1}}\right)
\ee
The norm of the vector $|Z_t'\rangle$ is then computed
as
\be
{\cal N} = \vec{X}_t'\cdot \vec{X}_t'-\vec{Y}_t' \cdot \vec{Y}_t'
\ee
The norm can be negative, and the definition of the new vector 
$|Z_{n+1}\rangle$ depends on the sign. 
In fact,
because we are actually  
doing {\it block}-Lanczos, implicitly operating not only on the 
vector$(X,Y)$ but also its RPA conjugate $(Y,X)$ simultaneously, 
there is a degree of freedom, corresponding to a hyperbolic 
rotation, in choosing the new vector.  The simplest 
choice for the vectors and corresponding 
RPA matrix elements is 
\be
|Z_{n+1}\rangle={1\over\sqrt{\cal N}}\left(\matrix{\vec{X}_t'\cr 
\vec{Y}_t'}\right),
\;\; a_{n+1} = \sqrt{\cal N},\;\; b_{n+1}=0;\;\; {\cal N}>0
\ee
and
\be
|Z_{n+1}\rangle={1\over\sqrt{\cal -N}}\left(\matrix{\vec{Y}_t'\cr 
\vec{X}_t'}\right),
\;\; a_{n+1} = 0,\;\; b_{n+1}=\sqrt{\cal -N},\;\; {\cal N} < 0
\ee
This completes the iteration cycle.

In analogy with the application to strength functions in the nuclear 
shell model, we start with  the vector given by
\be
|Z_1\rangle =\left(\matrix{ \vec{X}_1\cr \vec{Y}_1}\right)
=  \left(\matrix{ \vec{q}\cr 0}\right); 
\ee
With such a starting vector the algorithm 
manifestly preserves 
the energy-weighted sum rules:  
\begin{equation}
M_k= \sum_\nu \omega^k_\nu\left \langle \omega_\nu \right | \hat{Q} 
\left | 0 \right \rangle^2, \; k\; {\rm odd.}
\label{SumRule0}
\end{equation}
Using the eigenvector representation of the RPA matrix,
one can show \begin{equation} 
M_k= {1\over 2} ( \vec{q}, \vec{q} ) \left ( 
\matrix{ {\bf A} & {\bf B} \cr -{\bf B} & -{\bf A}
}\right )^k \left ( \matrix{ \vec{q}\cr  -\vec{q} } \right )
\label{SumRules}, \; k\; {\rm odd.}
\end{equation} 
With our method the $n$-th iterate respects
the odd-$k$ sum rules for $k\leq 2n-1$.

We now illustrate the method with a very simple model, 
a collective particle-hole
interaction fragmented by single-particle energies.  We 
consider states $i=1,...,N$ with matrix elements
$A_{ij} = \epsilon_i \delta_{ij} + \kappa q_i q_j$ 
and $B_{ij} = \kappa q_i q_j$.  Here $\epsilon$
represents the energy spacing of the particle-hole configurations,
$\kappa$ is the strength of the collective coupling to the 
field $Q$, and the components of the vector $q_i \propto 
i(N-i) \times r$, where the $r$ are Gaussian distributed 
random amplitudes, and normalize $|q|^2 = 1$.  The factor  
$i (N-i)$ weights the collective response towards the middle of 
the excitation spectrum. 
The parameter $\kappa$ should be positive for a repulsive collective
interaction such as the Coulomb that generates plasmons.  

In Fig. 1 we show the strength function for such an RPA matrix 
in a space of 500 states, with parameter values given in the caption.
The parameters were chosen to obtain moderate collectivity, with
a strong but broadly-fragmented collective excitation 
distributed over the spectrum. Fig.~1 also displays the $n$=3, 10,
and 50 approximants to the strength function,
where $n$ is the number of Lanczos 
vectors $|Z_i \rangle$, or, equivalently, the number of multiplications 
with the RPA matrix. 
 One sees that with a handful
of states, one state closely approximates the collective excitation and
the others distribute themselves over the remaining spectrum.
A better way to see the convergence of the strength function is
to plot its integral, $I(\omega) = \sum_\nu\Theta(\omega-\omega_\nu)
\langle \omega_\nu | Q | 0 \rangle^2$.  This is shown
in Fig. 2 for $n$=3 and 10.  After 50 iterations the integral of the 
strength function is virtually indistinguishable from the exact solution. 

We mention that the algorithm does not explicitly preserve the 
total strength $M_0$.  If there were no correlations in the 
ground state, that is, if the vectors $Y_i$ all vanished, then the 
total strength would be $|q|^2 =1$.  The non-trivial deviations from 1 
in our examples are related to the amount of correlations in the ground 
state. This is illustrated in Fig.~3, which is the integrated strength 
function for a model identical to that in Figs.~1,2 except that the 
collective interaction is attractive rather than repulsive. Here the total 
strength is about 3.7, i.e. quite different from 1. Fig.~3 illustrates 
how with $n=$3 and  10 the total approximate strength converges
rapidly to the exact value. (In the repulsive model of Fig.~2 the total 
strength had already converged by $n=3$.) 
Although we cannot prove this rapid convergence in all cases, 
it seems  likely in light of the strong constraints 
imposed by the odd-$k$ sum rules.

We anticipate that the algorithm will be particularly useful in
problems which require a single-particle dimensionality of the order
of tens or hundreds of thousands, but which allow a sparse
matrix approximation for the Hamiltonian, such as the local density
approximation.  This applies to molecular and condensed matter
physics modeled with the Kohn-Sham equations, and to nuclear
physics for excitations in deformed nuclei\cite{bo81}.  With the LDA Hamiltonian,
an efficient particle-hole representation can be constructed from
the orbital representation of holes and the coordinate-space representation
of particles\cite{ch94}.  The computational difficulty for the basic matrix-vector
multiplication then scales as the number of particles $N_p$ and the 
dimensionality of the single-particle space $M$ as  $M N_p^2\sim N_p^3$.  
Only a
fixed number of these operations, of the order of ten, are needed
to obtain the strength function to the accuracy of the fundamental
mean field approximation.  Thus the overall scaling of the method 
is $O(N_p^3)$.

     This study arose in the program at the Institute for Nuclear 
Theory,  ``Numerical methods for strongly interacting quantum
systems'', and we wish to thank J. Carlson and R. Wiringa for
providing that forum.  G.B. also thanks K. Yabana for many
discussions.  Financial support was provided by the INT under
Department of Energy Grant FG06-90ER40561 and by Department 
of Energy Grant DE-FG02-96ER40985.

\begin{figure}
\caption{Strength function for the model described in the text with 
$\kappa = 10$ and $\epsilon= 0.1$ (in arbitrary units) for 500 states, 
and the Lanczos approximants for 5, 10, and 50 Lanczos vectors.  
The scales for the abscissae are different because the strength is 
fragmented over a different number of states.}
\end{figure}

\begin{figure}
\caption{Integrated strength function for the model described in Fig. 1.
For 50 Lanczos vectors the integrated strength is virtually 
indistinguishable on this graph from the full calculation.
}
\end{figure}

\begin{figure}
\caption{The same as figure 2, except with the collective interaction 
is attractive, $\kappa=-10$. Notice that the 
total strength is not constrained, as described in the text, but 
has converged by the 10th iteration.
}
\end{figure}

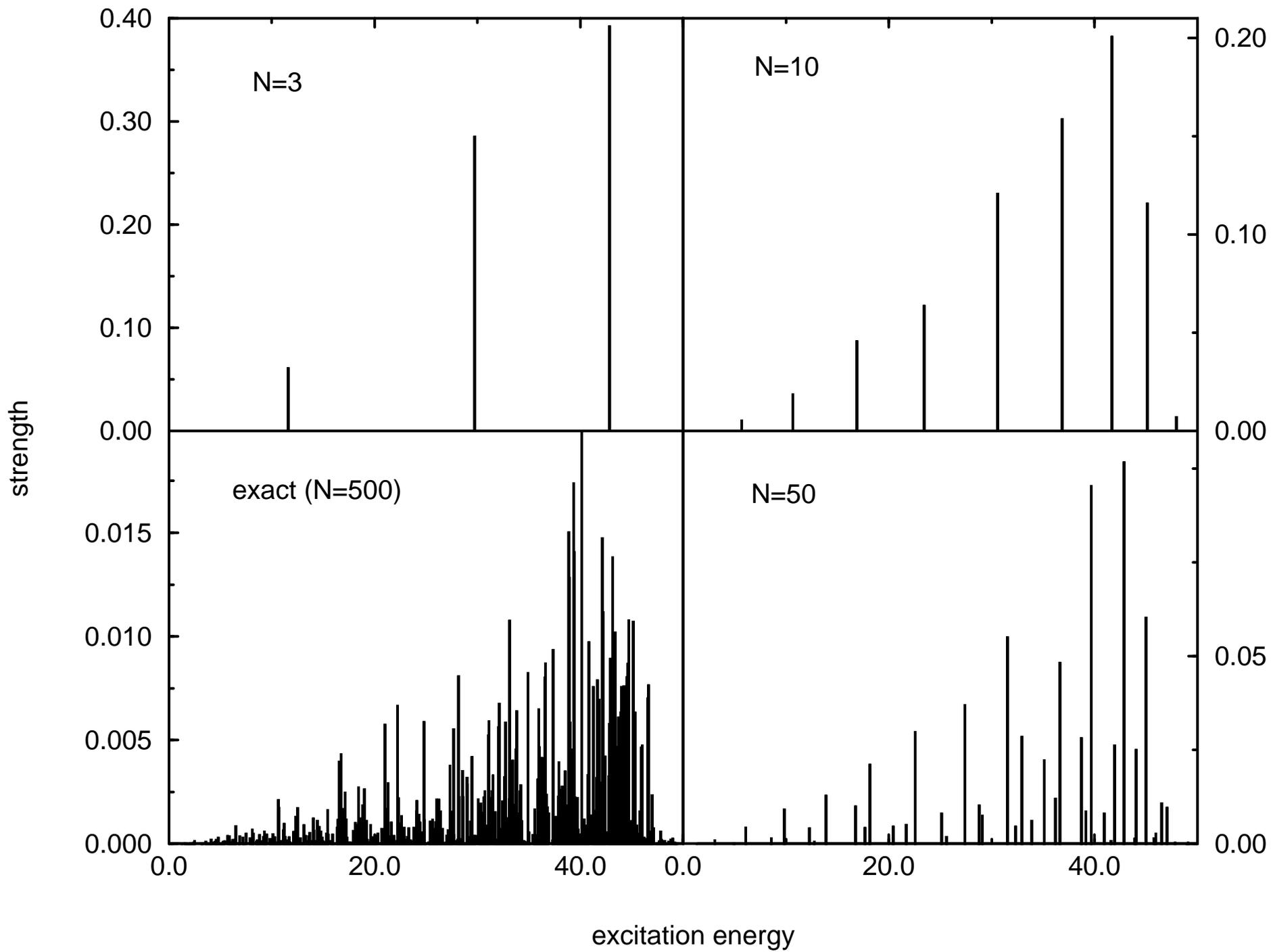
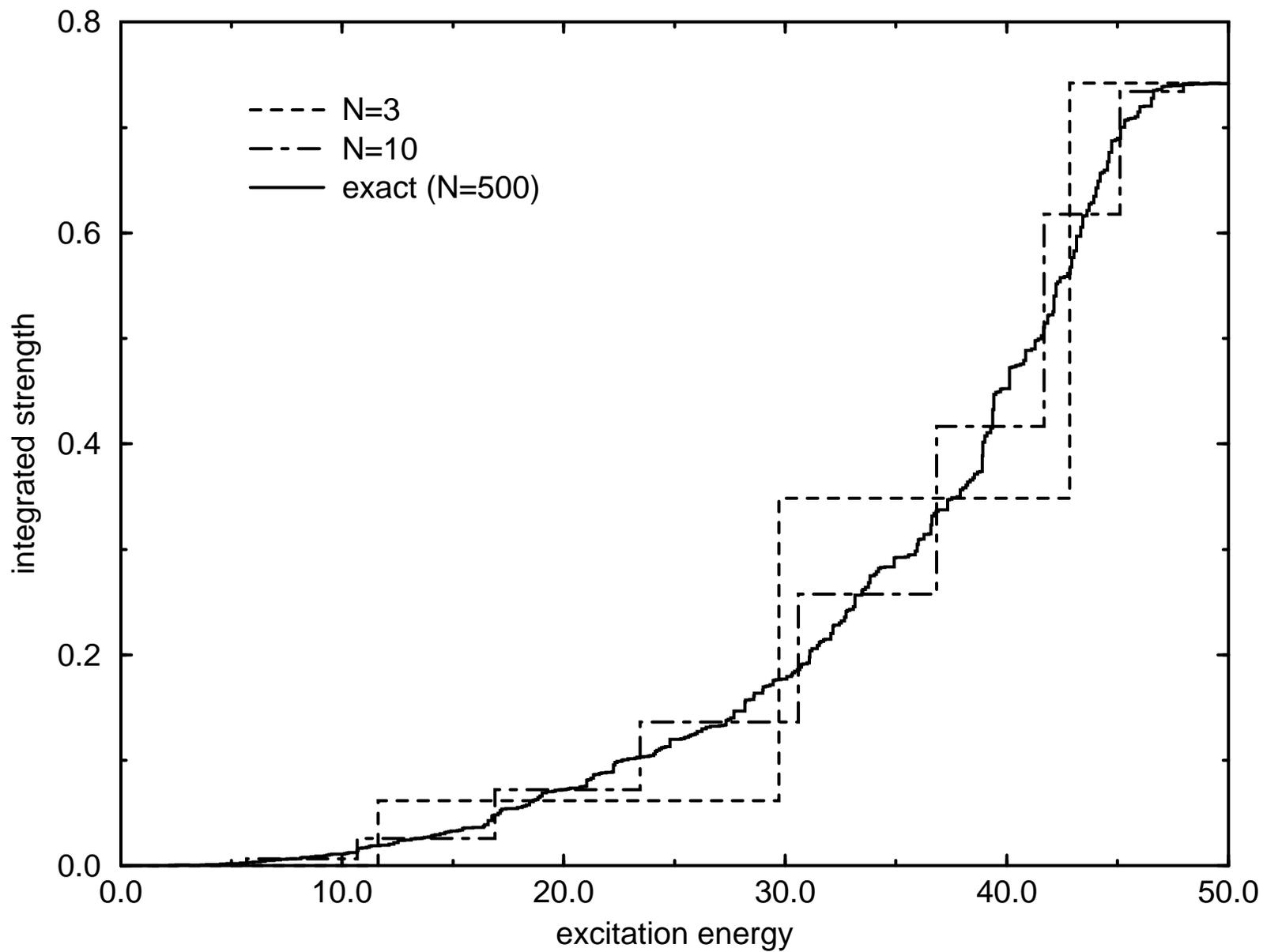
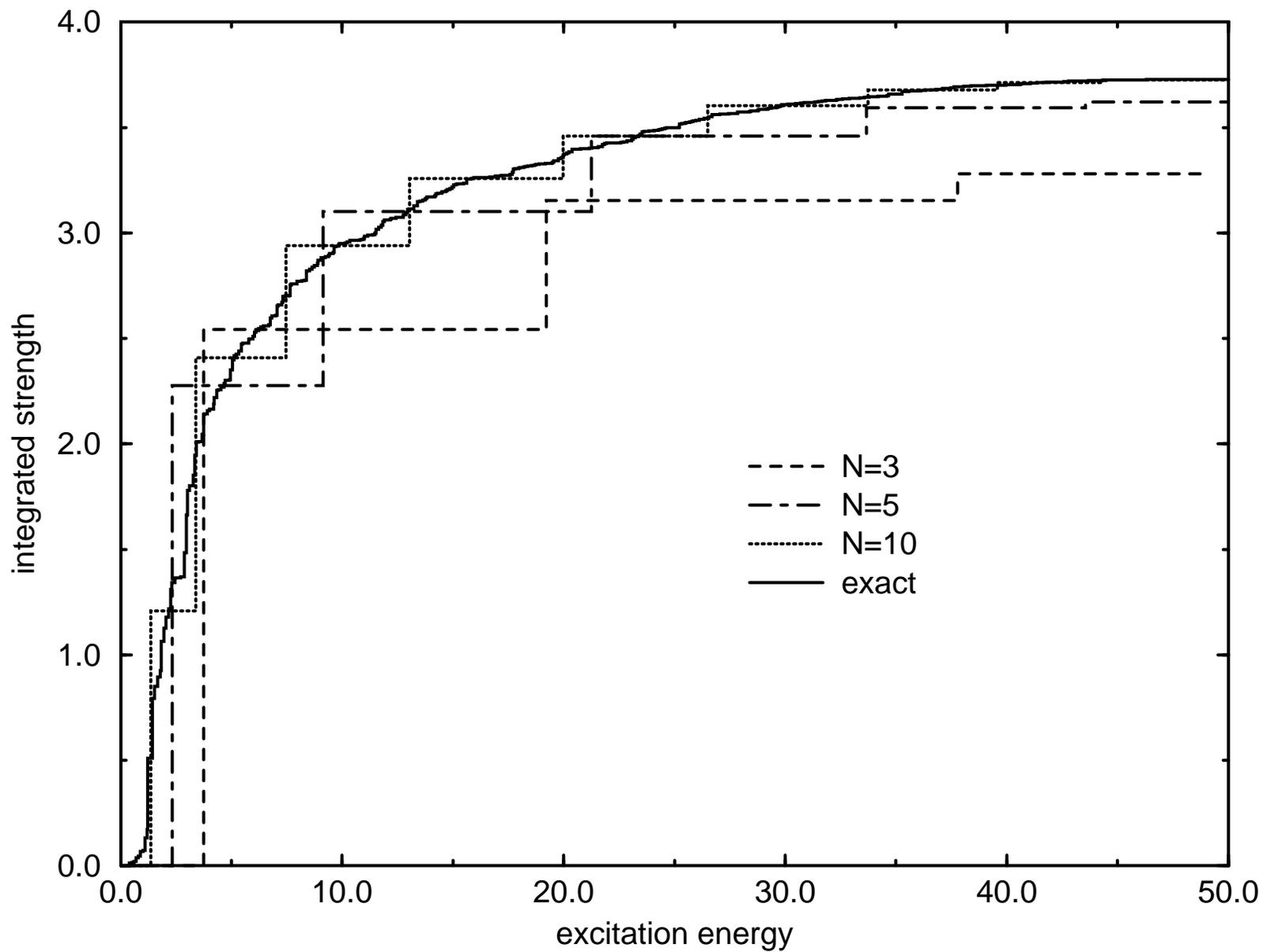
\end{document}